\documentclass[12pt]{iopart}
\usepackage{iopams}
\usepackage{graphicx}
\newcommand{\veps}{\varepsilon}

\begin{document}
\title{Limiting energy density and gravity in Riemann-Cartan space-time}

\author{A.V. Minkevich}

\address{Department of Theoretical Physics and Astrophysics, Belarusian State University, Minsk,
Belarus}
\address{Department of Mathematics and Informatics, Warmia and Mazury
University in Olsztyn, Poland} \eads{\mailto{minkav@bsu.by},
\mailto{awm@matman.uwm.edu.pl}}

\begin{abstract}
The gravitational interaction is discussed within the framework of gauge gravitational theory
in the Riemann-Cartan space-time. In the case of spatially homogeneous isotopic gravitating
systems the gravitational repulsion at extreme conditions near limiting energy density and the
transition from gravitational repulsion to attraction in dependence on energy density is
studied. The conversion to Friedman regime and the transition to gravitational repulsion at
very small energy densities is analyzed.

\end{abstract}
%\pacs{04.50.+h; 98.80.Cq; 11.15.-q; 95.36.+x}
% \submitto{\CQG}

\section{Introduction}

Within the framework of the general relativity theory  (GR), there are no restrictions for the
permissible values of the energy density for gravitating matter. Since in the case of ordinary
gravitating matter with positive energy density and non-negative pressure, the gravitational
interaction has the character of attraction increasing together with the energy density the
occurrence of singular states with divergent energy density which is unacceptable from a
physical point of view is inevitable in GR. In cosmology, this leads to cosmological problem
- the problem of the beginning of the Universe in time, inherent in the Big Bang model. Many
attempts have been made to solve this problem both within the framework of the metric
theory of gravitation and various alternative theories of gravity (see e.g. \cite{k1,k2}, review
\cite{k3}). However, most of the obtained partial regular solutions are unlikely to lead to the
solution of the cosmological problem, since at the same time there are singular cosmological
solutions that cannot be excluded from consideration for physical reasons. At the same time,
there is an approach in the relativistic theory of gravitation that allows us to solve the
cosmological problem due to the conclusion about the possible existence of a limiting (i.e.
maximum allowable) energy density in the nature. We are talking about the gauge theory of
gravity in Riemann-Cartan space-time (GTRC), known in the literature as Poincar\'{e} gauge
theory of gravity. The development of GTRC is connected with names T.W.B. Kibble, D.D.
Ivanenko, D.W. Sciama, A. Trautman and others and at present this theory is one of the
most important directions of the development of the theory of gravity (see for example
\cite{pbo}). GTRC is the theory based on generally accepted physical principles of classical
field theory and theory of fundamental physical interactions, including the principle of local
gauge invariance and it  is a direct generalization of the metric theory of gravity when the
group of tetrad Lorentz transformations is included in the gauge group corresponding to the
gravitational interaction. Investigations of GTRC based on general expression of gravitational
Lagrangian $\mathcal{L}_{\rm g}$ as function of gravitational field strengths - the curvature
$F^{ik}{}_{\mu\nu}$ and torsion $S^i{}_{\mu\nu}$  tensors including both the scalar
curvature and various invariants quadratic in the curvature and torsion tensors with indefinite
parameters assuming spatial parity, show that this theory by certain restrictions on indefinite
parameters satisfies the correspondence principle with GR in the case of gravitating systems
with energy densities much less than limiting energy density and allows to solve certain
principal problems of GR. \footnote{The most part of definitions and notations of our
previous work (see e.g. \cite{ m2}) are used below.} The conclusion about possible
existence of limiting energy density was obtained by investigation of  isotropic cosmology
built in the frame of GTRC (see e.g. \cite{m1,m2,m3,m4}). Gravitational interaction  at
extreme conditions (extremely high energy densities $\veps$ and pressures $p$) near limiting
energy density is repulsive excluding the appearance of singular state with divergent energy
density. The value of limiting energy density $\veps_{max}$ depends on content of
gravitating matter and indefinite parameters of gravitational Lagrangian $\mathcal{L}_{\rm
g}$: in the case of usual gravitating matter the dependence takes place on equation of state
$p=p(\veps)$; in the case of inflationary cosmological models the value $\veps_{max}$
depends also on parameters characterising scalar fields near limiting energy density.  Because
the gravitational interaction near limiting energy density is repulsive, the value $\veps_{max}$
should exceed the highest energy densities of existing astrophysical objects. The study of
gravitational interaction near limiting energy density can be interesting not only for cosmology
but also by investigation of superdense astrophysical objects.

The present paper is devoted to investigation of homogeneous isotropic gravitating systems
(HIGS) filled by gravitating matter with equation of state $p=p(\veps)$ in Riemann-Cartan
space-time at extreme conditions. At first the gravitational equations for HIGS are given.

\section{Equations for Homogeneous Isotropic Gravitating Systems in Riemann-Cartan Space-Time}

 Any HIGS in Riemann-Cartan space-time is described by three functions of time: the scale factor
of Robertson-Walker metric $R(t)$ and two torsion functions - scalar function $S_{1}(t)$
and pseudoscalar function $S_{2}(t)$. Gravitational equations for HIGS obtained in the
frame of GTRC take the following form \cite{m1, m6}
\begin{eqnarray}\label{2.1}%\fl
    \frac{k}{R^2} + (H-2S_1)^2 -S_2^2= \nonumber\\
    \frac{1}{{6f_0 Z}}
        \left[
            {\veps  -6 b S_2^2
            + \frac{\alpha }{4} \left( {\veps  - 3p - 12bS_2^2 } \right)^2 }
        \right],
\end{eqnarray}
\begin{eqnarray}\label{2.2}%\fl
    \dot{H}-2\dot{S}_1 +H (H-2S_1)= \nonumber\\
    -\frac{1} {{12f_0 Z}}
        \left[
            \veps  + 3p - \frac{\alpha } {2} \left( {\veps - 3p - 12bS_2^2 } \right)^2
        \right],
\end{eqnarray}
where $H=\dot{R}/R $  (a dot denotes the differentiation with respect to $x^0= c t$),
$k=+1,0,-1$ for closed, flat and open models respectively and $Z=1+\alpha\left( \veps - 3p -
12b S_2^2\right)$. The torsion functions $S_1$ and $S_2$ are:
\begin{eqnarray}\label{2.3}%\fl
    S_1  = -\frac{\alpha }{4Z} [\dot \veps
    - 3 \dot p + 12f_0 \omega H S_2^2
    -12( {2b - \omega f_0 } ) S_2 \dot S_2],
\end{eqnarray}
\begin{eqnarray}\label{2.4}
 S_{2}^{2}  = \frac{\veps - 3p}{12b} + \frac
{1-(b/2f_0) (1 +  \sqrt{X})} {12b \alpha (1- \omega/4)},
\end{eqnarray}
where $X=1+ \omega (f_0^2/b^2) [1- (b/f_0) - 2(1- \omega /4) \alpha ( \veps+ 3p)]\ge 0$ and
$f_0=\frac{c^{4}}{16\pi G}$ ($G$ is Newton's gravitational constant), $\alpha,  \omega, b$
are indefinite parameters.

 By using expressions for torsion functions and the equation
of the energy conservation law, which has the same form as in GR
\begin{equation}
\dot{\veps}+3H\left(\veps+ p\right)=0,
\end{equation}
we can express the Hubble parameter and its time derivative as functions of energy density
and pressure. Really by using (3) and (4) we find that
\begin{eqnarray}\label{35}
S_1  = -\frac{3f_0 \omega \alpha }{4bZ} HD ,
 \end{eqnarray}
where
\begin{eqnarray}\label{36}
D = \frac{1}{2} \left(3\frac{d p}{d \veps}-1 \right) \left(\veps+p\right)
 +\frac{1}{3}\left(\veps- 3p\right)
-\frac{b}{6f_0\alpha (1-\omega/4)} \sqrt{X}
\nonumber\\
+\frac{1-\omega (f_0/2b)}{2\sqrt{X}}\Big [\left(3\frac{d p}{d\veps}+1
\right) \left(\veps+p\right) \Big]
+ \frac{1-(b/2f_0)}{3\alpha (1-\omega/4)},
\end{eqnarray}
and $Z =  \frac{-\omega/4 + (b/2f_0)(1+ \sqrt{X})} {1-\omega/4}$. Then cosmological
equation (1) leads to the Hubble parameter in the form:
\begin{equation}
H=H_{\pm}= \pm \frac{ \sqrt{A_1}}{1+\frac{3f_0 \omega \alpha }{2bZ}D},
\end{equation}
where
\begin{eqnarray}
A_1=\frac{\veps- 3p}{12b} + \frac{\veps+ 3p}{12f_0 Z}
+ \frac{1-(b/2f_0)(1+ \sqrt{X})}{12b\alpha (1-\omega/4)}\left (1-\frac{b}{f_0 Z}\right)
\nonumber\\
+ \frac{(1-(b/2f_0)(1+ \sqrt{X}))^2}{24f_0 \alpha {Z} (1-\omega/4)^2} -  \frac{k}{R^2}.
\end{eqnarray}
By taking into account that $H-2S_1= H(1+\frac{3f_0 \omega \alpha }{2bZ}D)$ and
$\dot{D}=-3H(\veps+p)D_1$, where $D_1$ is certain function of energy density
\begin{eqnarray}
D_1= \frac{1}{6} \left(3\frac{d p}{d \veps}-1 \right) \left(3\frac{d p}{d \veps}+1 \right) + \frac{3}{2} \left(\veps+p \right) \frac{d^2p}{d^2\veps}
+\frac{f_0\omega}{6b \sqrt{X}}\left (1+3\frac{d p}{d \veps}\right)
\nonumber\\
 + \frac{1-\frac{\omega f_0}{2b}}{2\sqrt{X}}\Big [(1+\frac{d p}{d \veps})(1+3\frac{d p}{d \veps})
+3(\veps+p)\frac{d^2p}{d^2\veps}\Big]
\nonumber\\
+ \frac{1-\frac{\omega f_0}{2b}}{2X^{3/2}}\frac{f_0^2}{b^2} \omega \alpha (\veps+p) (1+3\frac{d p}{d \veps})^2 (1-\omega/4)
\end{eqnarray}
we obtain the time derivative of the Hubble parameter as functions of energy density
\begin{eqnarray}
\dot{H}= -H^2+\Big [A_2+\frac{9}{2} \frac{f_0\omega \alpha}{bZ} (\veps+p) H^2 \Big (D_1
\nonumber\\
+\frac{f_0\omega \alpha D}{2bZ \sqrt{X}}\left (1+3\frac{d p}{d \veps}\right ) \Big) \Big] \left [1+\frac{3f_0 \omega \alpha }{2bZ}D \right]^{-1},
\end{eqnarray}
where
\begin{equation}
A_2=-\frac{1}{12f_0Z}\left [\veps+3p-  \frac{(1-(b/2f_0) (1+\sqrt{X}) )^2}{2 \alpha (1-\omega/4)^2}\right]
\end{equation}
and $H$, $D$, $D_1$ are given by (7)-(10). With the purpose to analyse HIGS at extreme
conditions we transform obtained quantities and equations to dimensionless form by the
following way:
\begin{eqnarray}\label{13}%\fl
    x^0\to\tilde{x^0}=x^0/\sqrt{6f_0 \omega \alpha},& {}
    & H\to\tilde{H}=H\sqrt{6f_0 \omega \alpha},
    \nonumber\\
 \veps\to\tilde{\veps}=\omega \alpha\, \veps, & &  p\to\tilde{p}=\omega \alpha\,p,
      \nonumber\\
 S_{1,2}\to\tilde{S}_{1,2}=S_{1,2}\sqrt{6f_0 \omega \alpha}, & & b\to\tilde{b} = b/f_0,
     \nonumber\\
   R\to\tilde{R}=R/\sqrt{6f_0 \omega \alpha},  & &  \tilde{\veps}'+3\tilde{H}\left(\tilde{\veps}+\tilde{p}
\right)=0,
\end{eqnarray}
where prim denotes the differentiation with respect to $\tilde{x^0}$.  \footnote{It should be
noted that the scale factor $R(t)$ of Robertson-Walker metric is dimensionless and the
transformation  of $R$ given above is connected with the fact that the coefficient $k$ in term
$\frac{k}{R^2}$ does not change by considered transformation.} The quantities $X$ and
$Z$ are dimensionless and can be written in the form:
\begin{eqnarray}\label{17}
X=1+ \frac {\omega} {\tilde {b}} (\frac {1} {\tilde {b}} -1) - 2(1- \omega /4)
\frac {1} {\tilde {b}^2} (\tilde {\veps} + 3\tilde {p}),
\nonumber\\
Z =  \frac{-\omega/4 + (\tilde {b}/2)(1+ \sqrt{X})} {1-\omega/4}.
\end{eqnarray}
By using dimensionless value $\tilde {D}=\omega \alpha D$ we can write dimensionless
Hubble parameter in the form
\begin{equation}
\tilde {H}=\tilde{H}_{\pm}= \pm \frac{ \sqrt{\tilde{A_1}}}{1+\frac{3}{2\tilde{b}Z}\tilde{D}},
\end{equation}
where
\begin{eqnarray}
\tilde {A_1}=\frac{\tilde {\veps}- 3\tilde {p}}{2\tilde {b}} + \frac{\tilde {\veps}+ 3\tilde {p}}{2 Z}
+\omega \frac{1-(\tilde {b}/2)(1+ \sqrt{X})}{2\tilde {b} (1-\omega/4)}\left(1-\frac{\tilde {b}}{Z}\right)
\nonumber\\
+\omega \frac{(1-(\tilde {b}/2)(1+ \sqrt{X}))^2}{4{Z} (1-\omega/4)^2} -  \frac{k}{\tilde {R}^2}.
\end{eqnarray}
By taking into account that the quantity $D_1$ is dimensionless we write the time derivative
of the Hubble parameter (11) in the following dimensionless form:
\begin{eqnarray}
 \tilde {H}'= - \tilde {H}^2+\Big [ \tilde {A}_2+\frac{9}{2 \tilde{b}Z} ( \tilde {\veps}+ \tilde {p})  \tilde {H}^2 \Big (D_1
\nonumber\\
+\frac{ \tilde {D}}{2 \tilde {b}Z \sqrt{X}}\left (1+3\frac{d \tilde{ p} }{d \tilde { \veps} }\right ) \Big ) \Big] \left [1+\frac{3 \tilde{D} }{2 \tilde {b}Z}\right]^{-1},
\end{eqnarray}
where
\begin{equation}
 \tilde {A}_2=-\frac{1}{2Z}\left [ \tilde {\veps}+3 \tilde {p}- \omega \frac{(1-( \tilde {b}/2) (1+ \sqrt{X}))^2}{2  (1-\omega/4)^2} \right].
\end{equation}
By using obtained equations (15)-(18) we can investigate HIGS at extreme conditions near
limiting energy density. To do this, it is necessary to have the state equation of matter and the
values of indefinite parameters. As it was shown by investigation of isotropic cosmology in
the frame of GTRC, the most important results can be obtained by the following restrictions
on indefinite parameters:
\begin{equation}
0<1-\frac{b}{f_0}\ll 1,   \qquad 0 < \omega\ll 1,
\end{equation}
 and the value of parameter $\alpha^{-1}$ corresponds to some high
energy density, by which at cosmological asymptotics $\alpha \veps \ll 1$  By these
restrictions the value of limiting energy density $\veps_{max}$  is of order $(\omega
\alpha)^{-1} $.

 \section{Homogeneous Isotropic Gravitating Systems at extreme conditions}

Now we will investigate spatially flat HIGS by using equation of state for ultrarelativistic
matter $\tilde {p}=\tilde {\veps}/3$. By using restrictions (19) for parameters $b$ and
$\omega$ we obtain the following approximative expressions for $\tilde {H}$ and $\tilde
{H}'$ :
\begin{equation}
\tilde {H}=\tilde{H}_{\pm}= \pm \frac{ \sqrt{{2 \tilde {\veps} X (1+ \sqrt {X})}}}{4\tilde {\veps} + \sqrt {X} (1+\sqrt {X})},
\end{equation}
\begin{equation}
 \tilde {H}' + \tilde {H}^{2}= - \frac{2 \tilde {\veps}\sqrt {X}} {4\tilde {\veps} + \sqrt {X} (1+\sqrt {X})}
+ \frac{64 \tilde{\veps}^2 X^{3/2} (1+ \sqrt {X})} {[4\tilde {\veps} + \sqrt {X} (1+\sqrt {X})]^3}
\left [\frac{\tilde {\veps}}{X^{3/2}}+\frac{\tilde {\veps}}{X (1+\sqrt {X})}+\frac{1}{2\sqrt {X}}\right]
\end{equation}
where $ X=1-4 \tilde{\veps}$. Then the value of limiting energy density following from
$X=0$ is equal to $\tilde{\veps}_{max}=0,25$ and we have from (20)-(21) at the first
approximation with respect to $\sqrt {X}$ near the limiting energy density:
\begin{equation}
 \tilde {H}=\pm \frac{\sqrt{2 X}} {2},   \qquad  \tilde {H}'= 1- (3/2) \sqrt {X}
 \end{equation}
 what corresponds to the previously received estimate \cite{m7}. In the case under consideration ($\veps R^4=const$)  the implementation
  of the scale factor $R(t)$ can be represented near $\tilde{\veps}_{max}$ as:  $R(t)= c_1 exp{(c_2 t^2)}$ ($c_1$ and $c_2$ are constant).
The state with limiting energy density is achieved in the process of compression of HIGS
($\tilde{H}_{-}$)  being the initial state in the expansion process ($\tilde{H}_{+}$). The
results of numerical analysis of expressions (20)-(21) near limiting energy density are
presented in Fig.1 - Fig 3. \footnote{Fig.1 and Fig.2 do not reflect the behavior of the
corresponding quantities in the region of small values $\tilde {\veps} \rightarrow {0}$, while
the formulas are subject to correction. Numerical data are given with an accuracy of 0.001.}

  It follows from Fig.1 that the parameter $\tilde{H}_{+}$ ($\tilde{H}_{-}$)
 vanishing at the limiting energy density $\tilde{\veps}_{max}$ reaches its maximum (minimum)
 value $\tilde{H}_{+}= 0.272$ ($\tilde{H}_{-}= - 0.272$) at $\tilde{\veps}_1= 0.57  \tilde{\veps}_{max}$ .

 \begin{figure}
\centering
\includegraphics[scale=1.1]{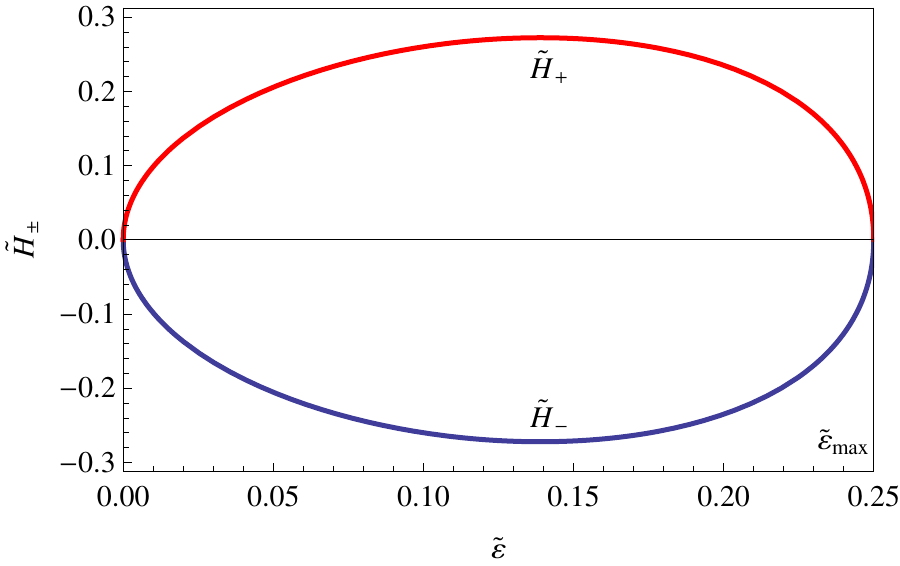}
\caption{Parameter $\tilde{H}=\tilde{H}_\pm$ as function of $\tilde{\veps}$ near limiting
energy density: $\tilde{H}_+$ (red line), $\tilde{H}_-$ (blue line)}\label{fig:01}
\end{figure}

\begin{figure}
\centering
\includegraphics[scale=1.1]{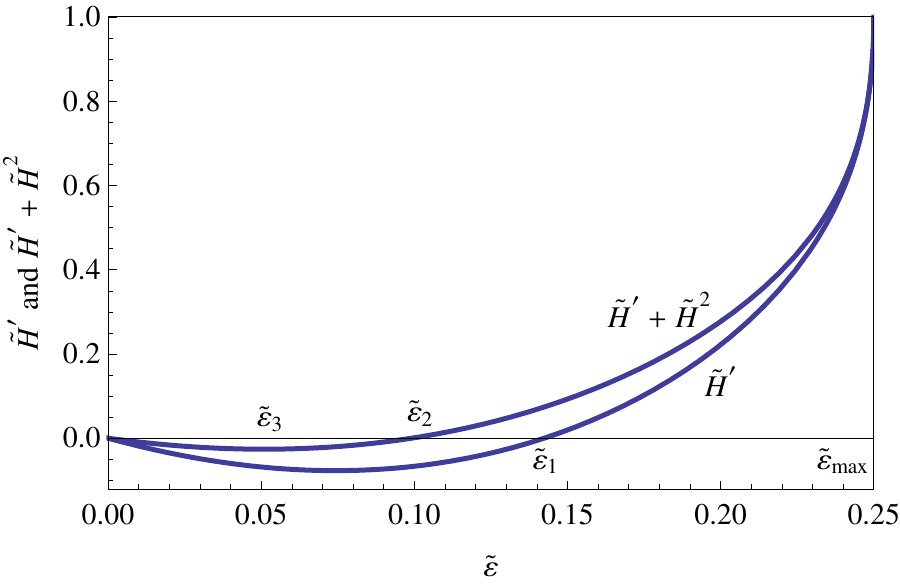}
\caption{Time derivative $\tilde{H}'$ and acceleration parameter $\tilde{H}' + \tilde{H}^2$ as functions of $\tilde{\veps}$ near limiting energy density} \label{fig:02}
\end{figure}

\begin{figure}
\centering
\includegraphics[scale=1.1]{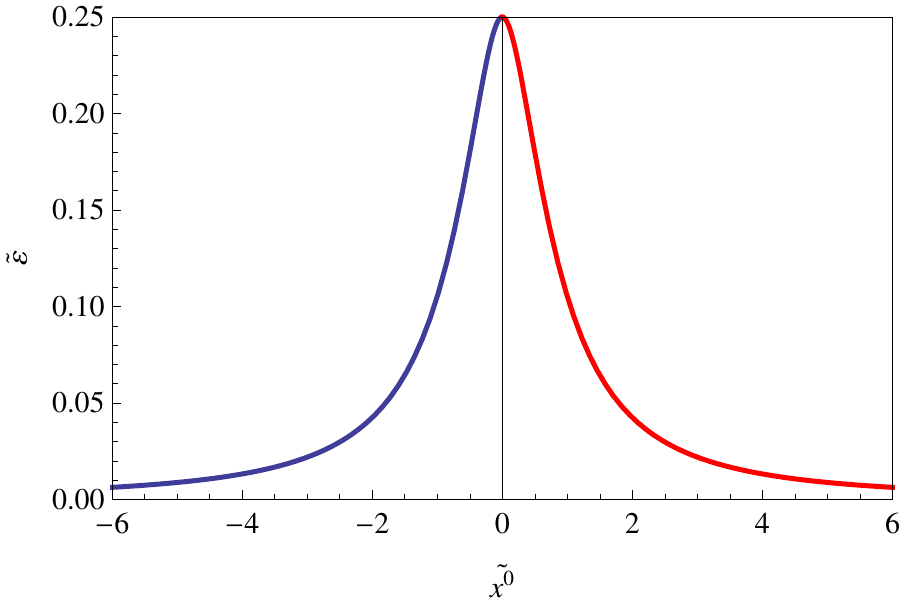}
\caption{Energy density $\tilde{\veps}$ as function of time $\tilde {x^{0}}$ near limiting energy density}
\label{fig:03}
\end{figure}

 As follows from Fig. 2 \footnote{The graphs in Fig. 2 are valid for both the $\tilde{H}_-$-solution and the $\tilde{H}_+$-solution.}the derivative $ \tilde {H}'$
  decreases from its maximum value $1$ to zero at $\tilde{\veps}_1$. The acceleration parameter
  $\tilde {H}' + \tilde {H}^{2}$ is
  also reduced and vanishes at the energy density $\tilde{\veps}_2 = 0.397\, \tilde{\veps}_{max} < \tilde{\veps}_1$. In the interval for the energy density
   ($\tilde{\veps}_2$, $\tilde{\veps}_{max}$), the gravitational
  interaction has the character of repulsion, and at the density $\tilde{\veps}_2$ there is a transition from gravitational repulsion to attraction.
  With a further decrease in the energy density, the negative acceleration parameter reaches its minimum value $\tilde {H}' + \tilde {H}^{2}=-0.026$
  at $\tilde{\veps}_3 = 0.204\, \tilde{\veps}_{max}$ corresponding to the maximum
  gravitational attraction force, which, as it decreases, approaches the gravitational attraction force of GR.
  The transition to the Friedman mode occurs when the value $\tilde{\veps}$ becomes much less than ${\tilde{\veps}}_{max}$ and the value X is approaching 1;
  then according to (20) $\tilde{H} \sim \sqrt {\tilde{\veps}}$, approximately such a transition occurs
  when $\tilde{\veps} =\tilde{\veps}_4 \sim 0.001\, \tilde{\veps}_{max}$. By using the equation of energy conservation in dimensional
  form (see (13)) we obtain the dependence $\tilde{\veps} =\tilde{\veps}(\tilde {x^{0}})$   at extreme conditions presented in Fig.3. Assuming
  that limiting energy density corresponds
   to $\tilde {x^{0}}=0$, we
   find an estimate for the moments of time $\tilde {x^{0}}_1=  \pm 0.718$, $\tilde {x^{0}}_2=  \pm 1.056$, $\tilde {x^{0}}_3=  \pm 1.768$,
   $\tilde {x^{0}}_4=  \pm 31.344$
   corresponding to $\tilde{\veps}_1$,
   $\tilde{\veps}_2$, $\tilde{\veps}_3$. $\tilde{\veps}_4$.
 By using obtained data we will estimate the time interval $\Delta {t}=\Delta {x^{0}/c}=
 ( \Delta {\tilde{x^{0}}/c}){ \sqrt{6f_0 \omega \alpha}}$  ($\Delta {\tilde {x^{0}}} = 2 {\tilde {x^{0}}_4}$) of transition
from the Friedman compression mode to the Friedman expansion mode.
 As noted earlier, the magnitude  $ \veps_{max}$
should exceed the energy density in the densest astrophysical objects and be less than the
Planck energy density. If we assume that the magnitude of the limiting density is two orders
of magnitude higher than the density of a neutron star $ ((\omega \alpha) \sim {10^{-37}}
(kg/m s^2)^{-1})$, then we find for the transition time from compression to expansion the
following estimation $\Delta {t}\approx 0.8 \cdot 10^{-3} s$ . With an increase in the limiting
energy density up to the Planck energy density ($ (\omega \alpha) \sim {10^{-113}} (kg/m
s^2)^{-1}$), the transition time is reduced to a time interval exceeding the Planck time
$t_{P}=10^{-43} s$ by one or two orders of magnitude.  It should be noted that the order
of the estimates obtained does not depend practically on the equation of state of matter ($0
\leq p \leq \veps/3$). After switching to the Friedman regime, the further evolution of HIGS
when taking into account changes in the equation of state of matter will correspond to GR
until the energy density decreases to values corresponding to the effective cosmological
constant $\Lambda=\frac{(1 - \frac{b}{f_0})^2} {6b\alpha}$ appearing in cosmological
equations in asymptotics that take the form of Friedman equations with $\Lambda$ \cite{m5,
m6}:
\begin{equation}\label{2.11}
    \frac{k}{R^2 } + H^2  = \frac{1}{6b }\left[\veps + \frac{1}{4\alpha} \left(1 - \frac{b}{f_0}\right)^2
     \right],
\end{equation}
\begin{equation}\label{2.12}
    \dot H + H^2  =  - \frac{1} {{12b }}\left[ (\veps + 3p) - \frac{1}{2\alpha}
    \left(1 - \frac{b}{f_0}\right)^2 \right].
\end{equation}

With the decrease $\veps$ and reversal of the acceleration parameter to zero, the gravitational
interaction becomes repulsive and the transition to accelerating mode occurs, The effective
cosmological constant in (23)-(24) induced by the vacuum torsion $S_{2}^{2(vac)} = \frac
{1  - b/f_0} {12b \alpha}$ leads to the change of gravitational interaction, when energy
density is small - the vacuum gravitational repulsion effect, which leads to accelerating
cosmological expansion \cite{m7}.

 If the limiting energy density exists in the nature, this should lead to
important physical consequences in astrophysics. First of all, we note that within the
framework of GTRC the stable astrophysical objects can exist at energy densities
significantly lower than the limiting density, at which the gravitational interaction has the
character of attraction. The properties of dense astrophysical objects with energy densities
comparable to the limiting energy density differ from what GR gives. The fundamental
consequence from a physical point of view is to prevent collapse and exclude singular states
with divergent energy density. Significant changes in the gravitational interaction in the case of
astrophysical objects with energy densities small compared to the limiting energy density take
place when their rotational moments interacting with torsion are taken into account. Thus, the
interaction of vacuum torsion with the rotational moments of astrophysical objects (stars,
galaxies) studied in the frame of minimum GTRC \cite{m8} leads to the appearance, in
addition to the Newtonian gravitational attraction force, of an additional force caused by their
interaction \cite{m7, m9, m10}.The magnitude of this force, as well as in general the physical
consequences associated with the interaction of torsion with the rotational moments of
astrophysical objects, depend on the restrictions imposed on the parameters b and $\alpha$.
Taking into account the value of the cosmological constant, following from observations, we
obtain a restriction on the parameters b and $\alpha$ while the condition
$0<1-\frac{b}{f_0}\ll 1$ takes place.  As a result, only one of the parameters $\alpha$ and b,
let's say parameter $\alpha$, is undefined. As for the parameter $\omega$, which is important
in the case of systems in extreme conditions and does not manifest itself in asymptotics, the
restriction $0 < \omega\ll 1$ was used in this paper. It should be noted that, in principle, a
theory with a limiting energy density can also be constructed in the case of $\omega \sim 1$.
It becomes important to find the values of parameters $\alpha$ and $\omega$ at which
GTRC agrees with the observational data.

\section{Conclusion}

Gauge gravitation theory in Riemann-Cartan space-time with a limiting energy density leads to
fundamental physical consequences that make important changes in existing ideas about the
world around us. Changing the gravitational interaction under extreme conditions near the
limiting energy density makes it possible to solve the problem of cosmological singularity
within the framework of classical theory without using quantum notions. The possible
existence of a limiting energy density leads to significant changes in the description of dense
astrophysical objects, usually identified with black holes. If the conclusion about the
existence of a limiting energy density turns out to be true, another constant will be added to
the number of fundamental physical constants, which determines the value of the limiting
energy density. Within the framework the considered theory, it will be probably a constant
$\alpha$.

\section{Acknowledgments}

The author is grateful to his colleagues who participated in the discussion of the studied
problems.
This article is dedicated to the memory of the author's son Arseny.

\end{document}